# Synergic mechanism and fabrication target for bipedal nano-motors

*Zhisong Wang*[*]

Institute of Modern Physics, Applied Ion Beam Physics Laboratory, Fudan University, Shanghai 200433, China

CORRESPONDING AUTHOR

Zhisong Wang

Institute of Modern Physics, Fudan University

Han-Dan Road 220, 200433 Shanghai, China

Phone: (86) 21 55664592

Fax: (86) 21 65642787

Email: wangzs@fudan.edu.cn

ABSTRACT

Inspired by the discovery of dimeric motor proteins capable of undergoing transportation in living cells, significant efforts have been expended recently to the fabrication of track-walking nano-motors possessing two foot-like components that can each bind or detach from an array of anchorage groups on the track in response to local events of reagent consumption. The central problem in fabricating bipedal nano-motors is how the motor as a whole can gain the synergic capacity of directional track-walking, given the fact that each pedal component alone is often incapable of any directional drift. Implemented bipedal motors to date solve this thermodynamically intricate problem by an intuitive strategy that requires a hetero-pedal motor, multiple anchorage species for the track, and multiple reagent species for motor operation. Here we performed realistic molecular mechanics calculations on molecule-scale models to identify a detailed molecular mechanism by which motor-level directionality arises from a homo-pedal motor along a minimally heterogeneous track. Optimally, the operation may be reduced to a random supply of a single species of reagents to allow the motor's autonomous functioning. The mechanism suggests a new class of fabrication targets of drastically reduced system requirements. Intriguingly, a defect form of the mechanism falls into the realm of the well-known Brownian motor mechanism, yet distinct new features emerge from the normal working of the mechanism.

Recently, several breakthroughs have started to emerge in the fabrication of nano-motors capable of walking directionally along an extended track(1-5). Track-walking nano-motors will enable active transportation of molecular cargos to any target area reachable by a nanometer-wide track, and will potentially have numerous applications. While the possibility of making nano-transporters was first mentioned by Feynman(6) nearly half a century ago, the recent fabrication efforts have been largely inspired by discoveries of dimeric motor proteins responsible for intracellular transportation along cytoskeletal filaments(7). Thus, artificial nano-walkers reported thus far generally possess two foot-like components, which are able to consume fuel reagents, detach from and bind to the foothold or anchorage components of the track. These pedal components often respond only to local events of fuel consumption, and a single pedal component is normally unable to drift directionally along the track. How a nano-motor comprising two such incapable pedal components can gain the synergic capability of directional movement as a whole is the central issue in fabricating bipedal nano-transporters. This is also a thermodynamically intricate problem that may bear general relevance to driven nano-systems(8-14).

As far as a bipedal motor is concerned, the entire motor's directional movement along the track can be ensured by the coordinating detachment and binding events of the two pedal components according to their relative position with reference to the chosen direction. The man-made bipedal walkers reported to date(1-4) rely more or less on an intuitively simple strategy for implementing position-selective detachment. Namely, chemically different feet and multiple species of anchorages are used to ensure that chemically distinguishable foot-anchorage bindings form depending on their being in the leading or trailing position with respect to the chosen direction. Position-dependent detachment is then executed by sequentially administering multiple fuel species which each selectively breaks a foot-anchorage combination of a certain chemical identity(1-4). This mechanism for motor-level direction selection requires as a necessary condition: (i) a hetero-pedal motor, (ii) multiple species of anchorages in an ordered arrangement along the track, (iii) multiple species of fuel reagents for energy-consuming breaking (and forming in some cases) of different types of foot-anchorage binding combinations, (iv) temporally ordered consumption of different reagent species. These are close to the highest requirements for motor-track systems, largely because the thermodynamics of this direction-rectification mechanism is close to the most primitive level. In fact, nearly all of the implemented motor-track systems use highly versatile DNA molecules as basic building blocks, and their success relies critically on the wealth of naturally existing reagents which enable detachment and binding in various DNA-based foot-anchorage combinations(1-4). To date, it appears formidable to go beyond engineered biomolecular systems to implement the heavily heterogeneity-relying mechanism.

Presumably, a thermodynamically advanced mechanism for motor-level direction rectification will reduce the system requirements for motor-track fabrication. The Brownian motor mechanism(15, 16) is able to rectify directional flow of independent particles over a periodic, asymmetric potential field in a thermodynamically appealing fashion. In principle, a nano-walker may be made by connecting via a molecular link two identical particles, each of which already gain some capability of biased drift along a common potential field via the Brownian motor mechanism(17, 18). However, this mechanism offers little on how connecting two independently directionless pedal components can ever bring about the synergic capability of the entire motor's directional movement.

By realistic molecular-mechanical calculations and dynamical simulations, we have identified a detailed molecular mechanism for converting a pair of polymer-connected, separately-directionless pedal components into a directional nano-walker. The mechanism suggests well-defined fabrication targets for bipedal nano-walkers and tracks, which may use chemically identical pedal components and no more than two anchorage species, and optimally, can be operated by a single species of fuel reagents. This drastic reduction in system requirements will extend the pool of candidate molecules for nano-walker construction beyond engineered biomolecules(8-14). Notably, a defect form of the mechanism for motor-level directionality falls into the realm of the Brownian motor mechanism, but distinct new features emerge out of the normal working of the mechanism. Interestingly, motors operating by this mechanism bear some key characters reminiscent of a dimeric motor protein involved in long-range intracellular transportation.

## RESULTS

### Basic requirements for walker-track system

The walker-track system comprises a walker having a pair of identical pedal components connected by a polymer chain, and a track as an extended structure supporting a periodic array of a single species of anchorage groups. Without losing generality, we assume that the foot-anchorage binding forms spontaneously and the reverse detachment is enabled by external stimuli(1-4). There are two basic requirements for the walker-track system: (i) In the vicinity of a foot-anchorage binding combination, the adjacent polymer is partially straightened and aligned towards a unique end of the array of anchorages. And, the polymer alignment must be disrupted after the foot is detached from the track. The alignment-pointed direction will thereafter be referred to as the forward direction of the walker. (ii) The interactions enabling the alignment must also modulate the overall affinity of the foot with the track so that a technical means can be developed for discriminately detaching an alignment-associated foot rather than an alignment-free one.

### Re-ordering of motor-track binding configurations upon shortening of inter-pedal polymer

We carried out single-molecule mechanics calculations for all possible walker-track interacting configurations to identify a wealth of molecular-physical mechanisms inherent in the seemingly simplistic walker-track system. Fig. 1 presents the total energies for certain major walker-track binding configurations (illustrated by the insets) as a function of contour length of the feet-connecting polymer (i.e. its largest end-to-end distance). The configurational energies were calculated using the following parameters for the walker-track system at room temperature $T_0$ = 298 K. The polymer's persistence length is $l_p$ = 1 nm, while the aligned length is $l_A$ = 0.5 nm, and the anchorage period is $d$ = 4 nm. The foot-anchorage binding energy without polymer alignment is $U_B = -10k_BT_0$ ( $k_B$ is the Boltzmann constant). We assume that the polymer alignment occurs spontaneously and lowers the walker-track configurational energies by $U_A = -5k_BT_0$.

Binding of both feet to the track reduces the conformational entropy of the polymer chain and causes it to experience mechanical strain. The magnitude of the internal strain for each walker-track configuration depends on its geometric details (see Eq. 2 in METHOD section). As the polymer's contour length approaches the anchorage period, the internal strain is raised to differing extents for different double-feet binding configurations. Therefore, a shortening of the inter-pedal polymer will cause re-ordering of the configurational energies, as can be clearly seen in Fig. 1. The strain-induced configurational re-ordering plays a key role in our direction rectification mechanism for the bipedal walker.

**Motor-level direction rectification by biased Brownian motion**

It is easily perceived that the partial alignment of the polymer at a foot-anchorage combination will bias diffusion of the other mobile foot towards the direction pointed by the polymer alignment (i.e. the forward direction). The mobile foot approaches a nearby anchorage only when the inter-pedal polymer adopts a stretched conformation of low entropy via intra-chain diffusion. This effect causes a free-energy barrier for the mobile foot's diffusive binding to the track. Apparently, less polymer stretching is required for the binding of the mobile foot to the neighboring anchorage pointed by the polymer alignment than to the anchorage on the opposite side. Consequently, the free-energy barrier is lower for the forward binding than rearward binding. The barrier difference $\Delta F$ can be directly ascertained from the data presented in Fig. 1. The barriers largely determine the average times needed for the diffusing foot to search and find the anchorages. An approximate one-dimensional treatment using the first-passage time theory(19, 20) predicts that the ratio of the average time for finding the forward anchorage compared to finding the rear anchorage is approximately proportional to $\exp(\Delta F / k_B T)$. Consequently, the diffusing foot approaches more frequently the forward anchorage than the rear one when the other foot is bound to the track with polymer alignment. Thus, the directional alignment, which is an asymmetric pattern localized within an individual foot-anchorage combination, is amplified via intra-chain diffusion into a dynamic asymmetry over a much larger range covering three consecutive anchorages.

Does this component of the biased Brownian motion of individual feet necessarily result in directional movement of the entire motor? As seen in Fig. 1, when the polymer's contour length is longer than 1.5 times the anchorage period, the lowest-energy configuration for the walker-track is one in which both feet are bound to the track and accompanied by polymer alignment (i.e. state I in Fig. 1). The walker-track ground state is translationally symmetric in the sense that both chemically identical feet adopt the same physical and chemical state. Any technical means of energy supply that enables the detachment of one foot is able to detach the other too. The walker is thus incapable of continual running, because indiscriminate detachment will inevitably cause concomitant bipedal detachment to derail the entire walker from the track. Furthermore, this symmetric ground state has no contribution to direction selection on the motor level. However, the alignment-caused diffusional bias is able to facilitate a biased drift of the bipedal walker despite the symmetric ground state, provided that concomitant detachment of both feet can be suppressed (e.g. by keeping low the frequency of executing energy supply). After an event of energy consumption detaches one foot from the walker-track ground state, the mobile foot is always more likely to bind to the anchorage preceding the standing foot that remains bound to the track. Thus, the bipedal walker as a whole may move back and forth for individual steps, but after a long run, a net drift will be accumulated towards the direction pointed by the polymer alignment. This is a typical Brownian motor that makes directional movement in an average sense. Thus, the polymer length range over which the symmetric ground state occurs may be termed the Brownian bias regime. Fig.2 A illustrates the walker's movement in this regime.

**Motor-level direction rectification by asymmetric ground state**

The double-feet binding configurations II and III, in which the polymer alignment occurs only once, at either the rear foot or front foot, may be regarded as being inversely asymmetric. In the long-polymer extreme states II and III tend to be degenerate in energy, because a long, fully relaxed inter-pedal polymer then will experience little internal strain in both states. As seen in Fig. 1, this configurational degeneracy is removed for finite polymer lengths. As the polymer's contour length ($l$) approaches the anchorage period ($d$), the polymer's internal energy becomes high enough to offset the foot-track binding energies and a completely new configurational hierarchy occurs. Intriguingly, the results in Fig. 1 expose a narrow but finite window of $1.1d < l < 1.5d$ in which the translationally asymmetric configuration II becomes the lowest-energy state, and the inversely asymmetric configuration III and the symmetric configuration I (i.e. the ground state in the long-polymer regime) are both elevated steeply to energies far beyond those of the single-foot binding configurations and become virtually inaccessible states.

Such a unique walker-track configurational hierarchy provides a ground for rectifying directional movement of the homo-pedal walker as a whole by means of a single type of foot detachment. The motor-track ground state is now asymmetric in the sense that the polymer alignment occurs at the rear foot but not the front foot. When the operational means is applied to the walker-track ground state, only the rear foot is detached, since the operation is designed for disrupting an alignment-modulated foot-track binding. The operation excites the walker-track system to a single-foot binding configuration, which will eventually decay back to the ground state by re-binding of the diffusing foot to the track either at an anchorage ahead of the standing foot or at the previous anchorage behind the standing foot. In the case of forward binding, both feet automatically exchange states so that the formerly standing foot becomes readily detachable for another step. For rearward binding, the formerly mobile foot resumes its rear position as well as a detachable state. Therefore, by each execution of the operational means, the walker's center-of-mass either moves forward or stays, but never moves back. This is essentially a mechanism of direction locking. We thus term the polymer length range for an asymmetric ground state, the direction locking regime. Fig. 2 illustrates the walker's movement in this regime.

Under repeated executions of the operation, the derailment of the entire walker from the track is suppressed as long as the average time for a mobile foot's search-and-binding is shorter than the average time for stimulus-enabled detachment of a standing foot. The inter-pedal polymer restricts diffusion of the mobile foot to the vicinity of the standing foot, and the first-passage time theory(19, 20) predicts an average search-and-binding time well below microseconds for $l < 10$ nm under zero load. The energy consumption, and ensuing structural changes within the foot-track interface leading to the ultimate detachment generally take microseconds or longer, particularly when chemical fuels are used. Therefore, the motor will be able to walk continually in a hand-over-hand manner towards the direction pointed by the polymer alignment under diffusive supply of fuel reagents. In the case of the energy supply being controllable, e.g. timely delivery of laser pulses or electrical signals, keeping a low frequency for the energy supply will ensure a long consecutive run length for the walker.

**Thermodynamic considerations**

The direction rectification mechanism based on an asymmetric ground state is distinctly different from the Brownian motor mechanism operating on a symmetric ground state. The latter mechanism is of a probabilistic nature, and rectifies rather straightforwardly the biased diffusion of the individual feet into a directional drift of the bipedal walker as a whole. The former mechanism selects from the asymmetric pattern of the walker-track ground state a unique motor-level directionality, and locks the walker's movement into it in a deterministic manner. The biased diffusion is no longer a necessary requirement for inducing motor-level directionality. However, it will promote energy efficiency by reducing the rearward binding of a mobile foot. Furthermore, the mechanism of direction rectification by biased diffusion is unable to support the continual run of the walker in practical implementation, and can be regarded as a defect form of the rectification mechanism based on the ground-state asymmetry in the long-polymer regime.

Motors operating in the direction locking regime are also different from a bipedal motor in which each pedal component is already an independent Brownian motor capable of continual directional drift. Previous studies(17, 18) predicted an inchworm gait for the latter category of bipedal motors, whilst the former category moves in a hand-over-hand manner.

The onset of a unique asymmetric ground state for the motor-track system under internal strain amounts to a mechanically mediated breaking of symmetry. The rectification of a motor-level direction by such ground-state asymmetry is a fairly new concept as compared to previous theories(15-18, 20, 21) and experiments(1-5) concerning track-walking nano-motors.

**Confirmation by dynamical simulations**

Our dynamical simulations further verified the direction rectification mechanism identified for homo-pedal walkers. The simulations assumed a value of 5.4 nm for the polymer's contour length, which corresponds to the direction locking regime (all other parameters for the walker-track system being the same as for Fig. 1). We assumed the rate for polymer alignment to be 10 $ms^{-1}$, and considered a small rate of 0.1 $ms^{-1}$ for the reverse, spontaneous disruption of the alignment. A diffusion coefficient of $3.5 \times 10^3$ $nm^2$/ms was assumed in calculating the search-and-binding rates of the mobile foot. We considered the rate for executing the designed operation (i.e. rate for applying stimuli e.g. adding fuels) on the magnitude of tens per millisecond. For the simulation to be realistic, we assumed for each execution of the operation a less-than-unit probability of success (50% in the simulation), and took into account a small but finite probability for erroneous breaking of an alignment-free foot-track binding by the operation (1% in the simulation). The simulation results summarized in Fig. 3 show that the walker can make hundreds of consecutive steps against an opposing force of a few pNs, and the average velocity can be a few nanometers per millisecond. Such a performance of velocity and run length is on the same magnitude as a dimeric biomotor such as the conventional kinesin(22, 23). In fact, both kinesin and the proposed motor walk in a hand-over-hand manner(24). As the rate for executing the operation increases, the walker's velocity increases but its consecutive run length drops. Both the velocity and run length decrease as the load increases, because the opposing force effectively increases the barrier for forward search of a mobile foot.

**DISCUSSION**

**Motor fabrication by dimerization of identical monomeric modules**

A general strategy for implementing the proposed motor is to first seek identical molecular modules consisting of a single pedal component covalently linking a polymer chain, and then to dimerize two of such basic modules by forming a connection between their polymers. The major challenge lies in fabricating basic modules ("motor monomers") capable of local polymer alignment, which in turn modulate the foot-track binding to allow its selective breaking. The molecular capabilities to be sought, namely polymer alignment and alignment-gated detachment, are all limited within the vicinity of an individual foot-anchorage combination. The dimeric motor can then gain the capacity of directional movement by fine-tuning the length of the polymeric bridge between both feet. According to our calculations, the range of polymer lengths suitable for the walker widens roughly linearly as the aligned length increases. We note that the strategy of motor fabrication by dimerizing identical monomers is also adopted by nature for making biological motors.

Thus, the inter-pedal polymer can be made of a plain, flexible polymer molecule whose contour length and persistence length are two major properties to be considered in experimental implementation. It is also necessary to minimize the inter-pedal polymer's affinity with the track so as to minimize interference with the walker's movement. Candidate molecules for the inter-pedal polymer include short single-stranded DNA and peptides, whose nucleotide or amino acid sequences can be designed to meet the length and affinity requirements. We note that the use of DNA strands as the inter-pedal polymer has been reported before in DNA-based walkers(1-5). The track substrate can be a periodically extending structure made of e.g. rigid polymeric structures or functionalized nanotubes, to which the anchorage groups can be tethered, preferably by chemical bonds for stability. The track substrate must be rigid enough to avoid large-scale curving in aqueous environments, and this requires the substrate's overall persistence length to be much larger than the anchorage period. Candidate systems for the track substrate include DNA double helical structures as walked by previously implemented nucleotide-based motors(1-4), filamentous protein structures as walked by biological motors(7), and synthetic structures like properly functionalized carbon nanotubes. Another factor to be considered in the experimental implementation is the foot-anchorage binding energies, which not only determine the minimum amount of energy supply for detachment but also play a part in determining the values of the finely-tuned polymer length (see METHOD section). Implementation of the foot-anchorage bindings and the related unbinding operation will be discussed in the following subsection.

**Two specific fabrication targets**

Fig. 4 schematically illustrates two possible scenarios for implementing foot-anchorage binding and polymer alignment. In the first scenario, the alignment is caused by binding the polymer directly to a track component (a secondary anchorage) in the vicinity of a foot-hosting anchorage. It is required that some correlation forms between the polymer-track binding and the nearby foot-anchorage binding because of their close proximity. The operational means must then simultaneously disrupt the mutually affected foot-anchorage and polymer-track bindings without causing a severe destabilizing effect on an alignment-free foot-anchorage combination. In the second scenario, the alignment is caused by binding the polymer to an extended foot, which in turn binds with an anchorage in a fixed orientation. The polymer-foot binding is expected to cause a substantial structural adjustment within the foot. The detachment operation desirably affects only a structurally adjusted foot, and results in a reverse structural change to disfavor the polymer binding after foot detachment. Interestingly, a polymer alignment similar to the second scenario has been found to occur in the dimeric biomotor kinesin(25).

Both of the proposed fabrication targets are homo-pedal walkers. The first scenario requires two anchorage species, i.e. a foot-holding anchorage and an alignment-responsible secondary anchorage. The second scenario needs a single anchorage species to accommodate the directional binding of the pedal components. For both scenarios, a single type foot-anchorage binding combination needs to be broken by an exterior energy supply. If the foot-anchorage binding and polymer alignment occur spontaneously, the motor will be able to run autonomously under repeated execution of a single type of detachment operation. Therefore, both fabrication targets are already close to the minimum level of operational and system requirements for a bipedal nano-motor.

The foot component and anchorage groups for both fabrication targets may be implemented using a variety of molecular systems including DNA(1-5), peptides and synthetic molecules(8-14). Nearly all bipedal walkers implemented thus far(1-4) have used DNA strands to make the foot and anchorage groups. A pair of foot and anchorage strands is designed to have complementary nucleotide sequences, and upon diffusive encounter spontaneously form multiple hydrogen bonds with each other. Such a foot-anchorage connection is disrupted using either naturally existing enzyme proteins, which site-specifically cleave a DNA duplex(3, 4), or "unset" strands, which form a more stable connection with one of the foot and anchorage strands thereby setting the other free(1, 2). Short DNA strands appear to be suitable candidates for both the foot components and the anchorage groups (including the secondary anchorage group) for the first set of fabrication targets mentioned above. By means of pre-arranged sequence complementarity between these strands, both the foot-anchorage binding energy and the alignment energy can be designed. The detachment techniques used to operate the reported DNA-based walkers(1-4) can be adapted for operating the motor of current interest. In the second set of fabrication targets, the polymer alignment arranges the foot and the adjacent polymer into a hairpin-like configuration, which resembles the β sheet structure occurring abundantly in proteins. Thus, peptides appear to be suitable candidates for forming the foot component as well as the inter-pedal polymer for the fabrication targets of the second type. The foot-track binding can be substantiated by hydrogen bonds or electrostatic attraction occurring at the foot-track interface. Detachment is possible by stimuli-induced structural re-arrangements within the foot component, which either relocate the hydrogen bonding partners or change the charge distribution. For a peptide foot-polymer-foot walker, hydrogen bonding seems to be a particularly suitable means for polymer alignment because it is short-ranged and sensitive to local structural change. Finally, we note that binding and stimuli-caused unbinding have also been implemented in synthetic molecules like rotaxanes and catenanes(8-14). These synthetic systems and related techniques may in principle be adapted for construction and operation of the motor proposed here.

**A general strategy for implementing alignment-gated detachment**

Implementation of alignment-gated detachment is probably the most demanding issue in the fabrication of the homo-pedal motor. Notably, a quantitative analysis of the configurational hierarchy in Fig. 1 points to a convenient strategy for achieving alignment-gated detachment. Following the first scenario for polymer alignment, let us consider an operational means that is mainly designed to disrupt the polymer alignment instead of the foot-anchorage binding. Consider the instances where the removal of a polymer alignment leads to a transient weakening of a nearby foot-anchorage binding because of its proximity to the aligned part of the polymer. Suppose that the resulting instant energy rise ($\Delta E$) is not enough to compensate for the foot-anchorage binding energy ($U_B$), i.e. $\Delta E < |U_B|$. This sub-threshold weakening is unable to derail the motor from any single-foot binding configuration. Nevertheless, the above-designed operation can cause foot detachment from the walker-track ground state if $\Delta E$ falls within the range $|U_B| > \Delta E > E_C(VI) - E_C(IV)$. Here $E_C(VI)$ and $E_C(IV)$ are the energies for the highest-lying single-foot binding configuration and the alignment-free double-feet binding configuration (see insets in Fig. 1). The operation is then able to instantly excite the walker-track system to an energy beyond that for the highest-lying single-foot binding configuration. This transient high-energy state, in which the inter-pedal polymer is unsustainably overstretched, readily decays to a single-foot binding via spontaneous detachment of the formerly alignment-accompanying foot, because this foot's binding to the track is instantly weaker than that of the other undisturbed foot. Thus, an alignment-disrupting operation so designed can ensure alignment-gated detachment, and also suppress derailment. The above strategy for implementing discriminate detachment works ideally for those polymer lengths suitable for bipedal walking, because the biggest room for $\Delta E$ exists for these polymer lengths. For a longer polymer, $E_C(VI) - E_C(IV) \to |U_B|$, and the room for $\Delta E$ tends to vanish. The above strategy thus ceases to work for long polymers.

**CONCLUSIONS AND OUTLOOK**

How a nano-motor made of two locally responding and independently directionless pedal components can gain the synergic capacity of directional track-walking at the entire motor level is a technically important and thermodynamically intriguing problem. Bipedal nano-motors implemented to date have gained motor-level directionality by a thermodynamically primitive strategy, which requires a high level of heterogeneity of fabricated motor-track systems, and has been implementable thus far largely by virtue of using versatile biomolecules as building blocks and the rich biochemical mechanisms for motor operation. Here, we have established a conceptually new molecular mechanism for motor-level direction rectification that drastically reduces the requirements for fabrication and operation of the motor-track systems. The main findings are summarized below.

(i) A bipedal nano-motor can be obtained by linking via a polymer chain two identical synthetic pedal components which each spontaneously bind to the extended track and allow occurrence of a localized partial alignment of the polymer chain in the vicinity of the foot-track interface. Such a basic system of a "homo-dimer" as the motor plus a minimally heterogeneous track can gain the synergic capacity of motor-level directionality by fine-tuning the length of the inter-pedal polymer chain. The operational task is a single type of stimuli-powered discriminate detachment of a track-bound pedal component that is accompanied by a nearby polymeric alignment.

(ii) Fine-tuning the inter-pedal polymer length is a key factor in bringing about motor-level directionality. Molecular mechanics calculations revealed distinct regimes for the intrinsic mechanics of the basic motor-track system as the inter-pedal polymer is shortened. For the majority of polymer lengths, the motor-track is incapable of sustained directional translation. Only when the polymer length falls within a narrow but finite range, an asymmetric ground state occurs for motor-track binding configurations and other symmetrically complementary configurations are elevated to inaccessible energies. The onset of the unique asymmetric ground state amounts to a breaking of the symmetry, which in turn forms the basis for a motor-level direction rectification.

(iii) The discovered molecular mechanism of rectifying motor-level directionality from ground-state asymmetry suggests a new class of homo-pedal nano-motors which can be fabricated by dimerizing identical synthetic pedal components. The alignment-gated selective detachment being likely a demanding experimental issue, can be implemented via a rather general strategy by exploiting the unique motor-track mechanics in the regime of the asymmetric ground state. Two basic categories of synthesis targets for the motor-track were outlined, and these may be implemented using a variety of candidate systems from engineered biomolecules(1-5) to synthetic molecules(8-14). This study thus points to a systematic approach to developing a large class of new-generation bipedal nano-motors operating on the same core mechanism. If the single-type of exterior operation is executed by a single species of reagents, the homo-pedal motor will be able to function autonomously under random supply of the reagents.

The length-sensitive mechanism of rectifying motor-level directionality from ground-state asymmetry has intriguing thermodynamic implications. The mechanism possesses a feature of deterministic direction-locking. A defective form for an elongated polymer however falls within the realm of the well-studied Brownian motor mechanism. Furthermore, the bipedal motor operating on the identified mechanism shares some key characteristics with the dimeric biomotor kinesin, e.g. both contain a partial alignment of the inter-pedal polymer, and both walk in a hand-over-hand manner. The thermodynamic aspects and possible biological relevance of the identified mechanism, and above all its various experimental implementations, are worthy of future study.

**METHODS**

**Molecular mechanics calculation**

As illustrated by insets in Fig.1, a walker-track binding configuration is specified by the number of feet bound to the track ($n_B$, which may be $n_B = 0$, 1 or 2), the total number of occurrences of polymer alignment ($n_A$, being 0, 1 or 2), and the number of alignments at the leading foot ($n_L$, being 0 or 1). A subtlety to be considered is that the directional alignment at the leading foot forces the formation of a half loop in the polymer when both feet are bound to the track. The total configurational energy ($E_C$) is a sum of the Helmholtz free energy of the polymer chain ($F_H$), the foot-anchorage binding energy ($U_B$), and the energy change caused by polymer alignment ($U_A$),

$$E_C = F_H(n_B, n_A, n_L) + n_B U_B + n_A U_A. \quad (1)$$

Single-molecule experiments of mechanically stretched polymers (26, 27) have revealed that the measured force-extension curves are generally well described by an equation derived from the worm-like chain model(28) $f = (k_B T / l_p)[x/l_C + 1/4(1 - x/l_C)^2 - 1/4]$, where $x$ is the average end-to-end extension of a polymer chain in the direction of the applied force $f$, $l_C$ is the polymer's contour length, and $l_p$ is the persistence length that quantifies the polymer's bending rigidity. $k_B$ is the Boltzmann constant, and $T$ is the absolute temperature. This experimentally verified equation yields the Helmholtz free energy of the walker polymer as:

$$F_H(n_B, n_A, n_L) = (k_B T)\left(\frac{l_{eff}}{l_p}\right)\left[\frac{(d_{eff}/l_{eff})^2 (3 - 2 d_{eff}/l_{eff})}{4(1 - d_{eff}/l_{eff})}\right]. \quad (2)$$

Here, $l_{eff}$ is the effective polymer length defined for each walker-track configuration, and $d_{eff}$ is the translational distance spanned by the $l_{eff}$ portion of the polymer. A geometry analysis yields:

$$l_{eff} = l - n_A l_A - n_L l_L, \quad (3)$$

$$d_{eff} = d - (n_A - 2 n_L) a. \quad (4)$$

Here, $l$ is the polymer's contour length, $l_A$ is the part fixed by the alignment, $l_L$ is the contour length minimally required for forming a half loop. $d$ is the anchorage period, and $a$ is the projection of $l_A$ along the anchorage array (i.e. the track) (we assume $a \approx l_A$). Both polymer theories(29, 30) and experiments(31, 32) have revealed that the probability for a polymer chain to form a closed loop reaches a peak when the polymer's contour length is several times the persistence length, and decreases sharply when the contour length is reduced below 1.5 times the persistence length. We thus take $l_L = 0.75 l_p$.

**Kinetic Monte Carlo simulation**

Major processes occurring in the walker-track system include the foot detachment under external stimulus (e.g. fuel molecules), the diffusive search and re-binding of a mobile foot to a new anchorage, and polymer alignment. All of these processes are stochastic in nature, and can be quantified by rate constants. The track-walking dynamics can then be simulated using the kinetic Monte Carlo method(33, 34). The rate for the diffusive search and binding was calculated using the first passage time theory(19, 20) based on energy barriers derived from the configurational calculation. The detachment and alignment rates are simply assigned realistic values.

## ACKNOWLEDGMENTS

This work was partly funded by the National Natural Science Foundation of China (Grant No. 90403006), the Chinese Ministry of Education (Program for New Century Excellent Talents in University), the Shanghai Education Development Foundation (Shu-Guang Program), and the Shanghai Pu-Jiang Program.

## FIGURE CAPTIONS

**Figure 1.** Energies for motor-track binding configurations as a function of the inter-pedal polymer's contour length (i.e. its largest end-to-end distance). The insets illustrate the motor-track configurations for a specific scenario for implementing the motor. Because the polymer is mechanically strained to a differing degree in different motor-track bipedal binding configurations, the shortening of the polymer's contour length causes a re-ordering of the configurational energies. Consequently, two regimes for motor-track configurational hierarchy occur (marked by shadowed areas). Also indicated is the bias caused by the polymer alignment at a track-bound foot, i.e. the difference between free-energy barriers for forward and backward binding of the other mobile foot. The motor-track parameters are given in the main text.

**Figure 2.** Schematic illustration of the motor's movement under the designed operation. (A) Comparison between the direction locking regime and Brownian bias regime (both marked in Fig. 1). The molecular events of foot detachment by the designed operation, spontaneous polymer alignment and a mobile foot's diffusive binding to the track, are represented by filled black arrows, filled grey arrows and unfilled arrows, respectively. These processes cause transitions between the different motor-track binding states (marked as in Fig. 1). In the bias regime, state I is the lowest-energy state from which the operation will excite the walker-track to state V or cause derailment of the entire motor from the track to terminate any run. As for the next two low-lying states, II and III, the operation will selectively detach the trailing foot in the former state and the leading foot in the latter. Since state II is lower in energy, and thus occurs with a higher probability than state III, the motor's center-of-mass will move back and forth, but with a bias toward the forward direction if the terminating event of total derailment is suppressed. In the locking regime, states I and III become energetically inaccessible (as shown above the shadowed area). State II is left as the ground state, from which the operation will invariably detach the trailing foot. As a consequence, the motor's center-of-mass moves forward or stays, but never turns back. (B) Illustartion of the motor's movement cycle that ensures a full step forward in the locking regime.

**Figure 3.** Performance of the bipedal motor predicted by simulations using realistic parameters. The parameters are given in the main text. (A) and (B) show average velocity and average consecutive run length as a function of the rate for executing the designed operation under zero load. (C) and (D) show the velocity and run length as a function of opposing loads for a fixed operation rate of 1 $ms^{-1}$.

**Figure 4**. Schematic illustrations for two basic categories of synthesis targets for the bipedal motor and track. Each motor monomer contains a plain polymer chain (curve) linking a foot component (black symbols) and a component (open circles) responsible for dimerizing the motor monomers.

Figure 1

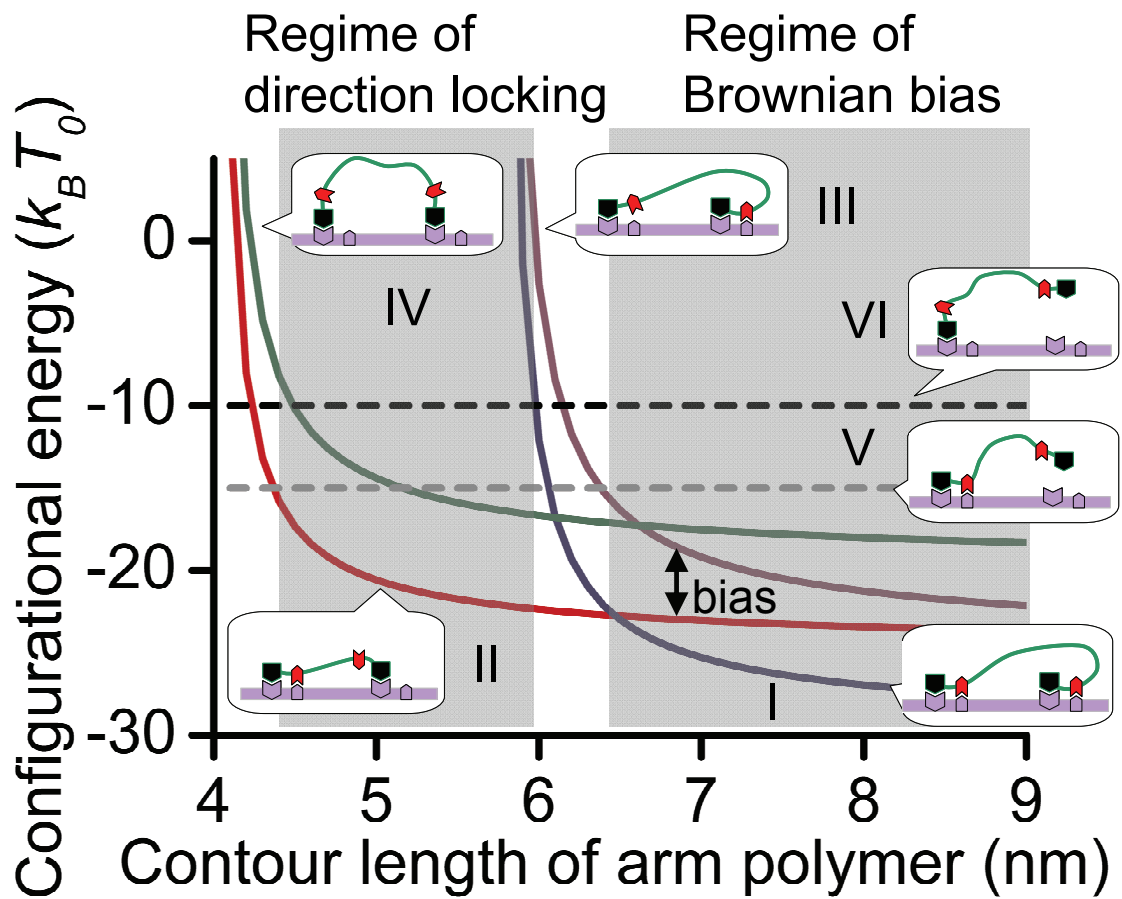

Figure 2

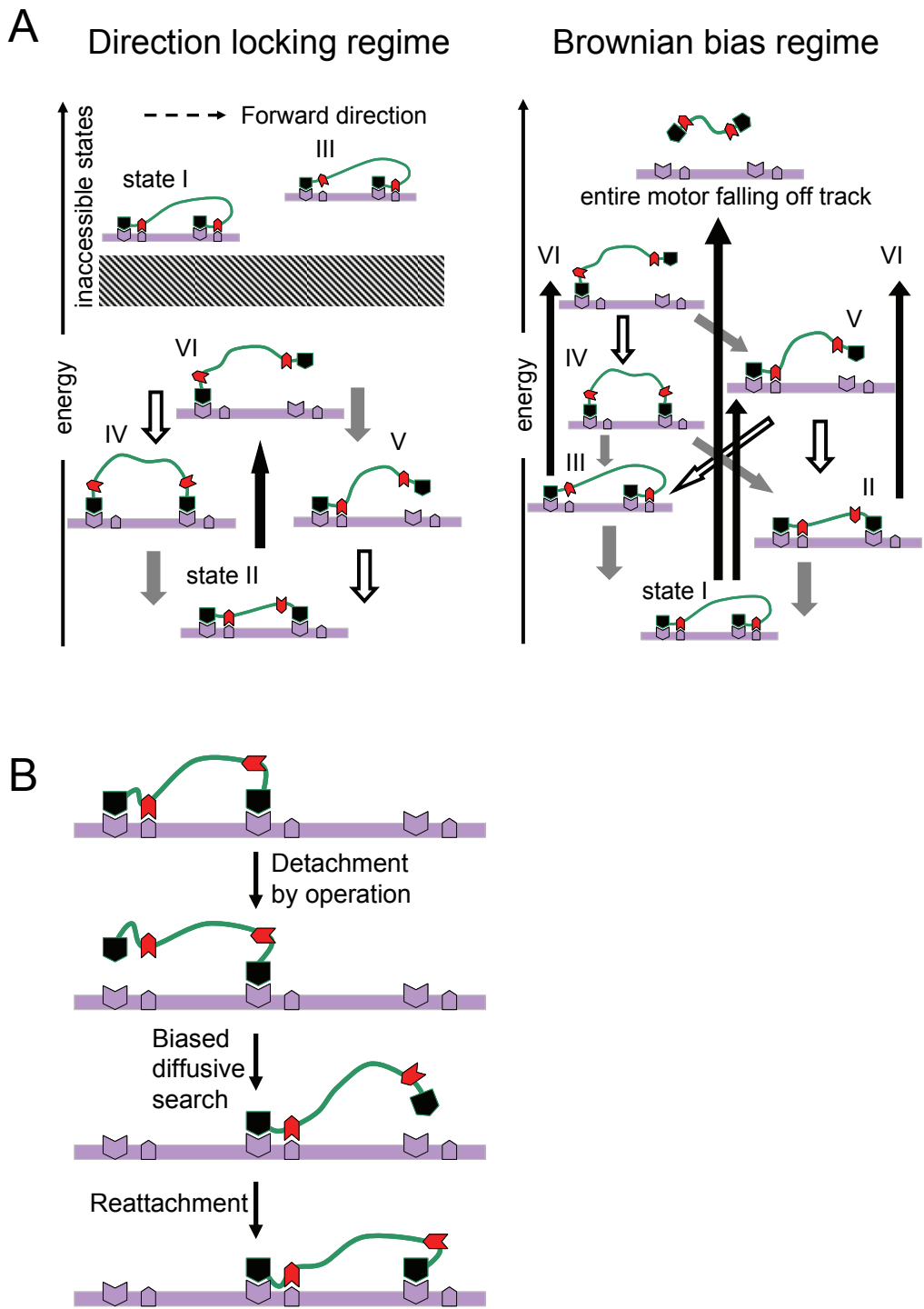

Figure 3

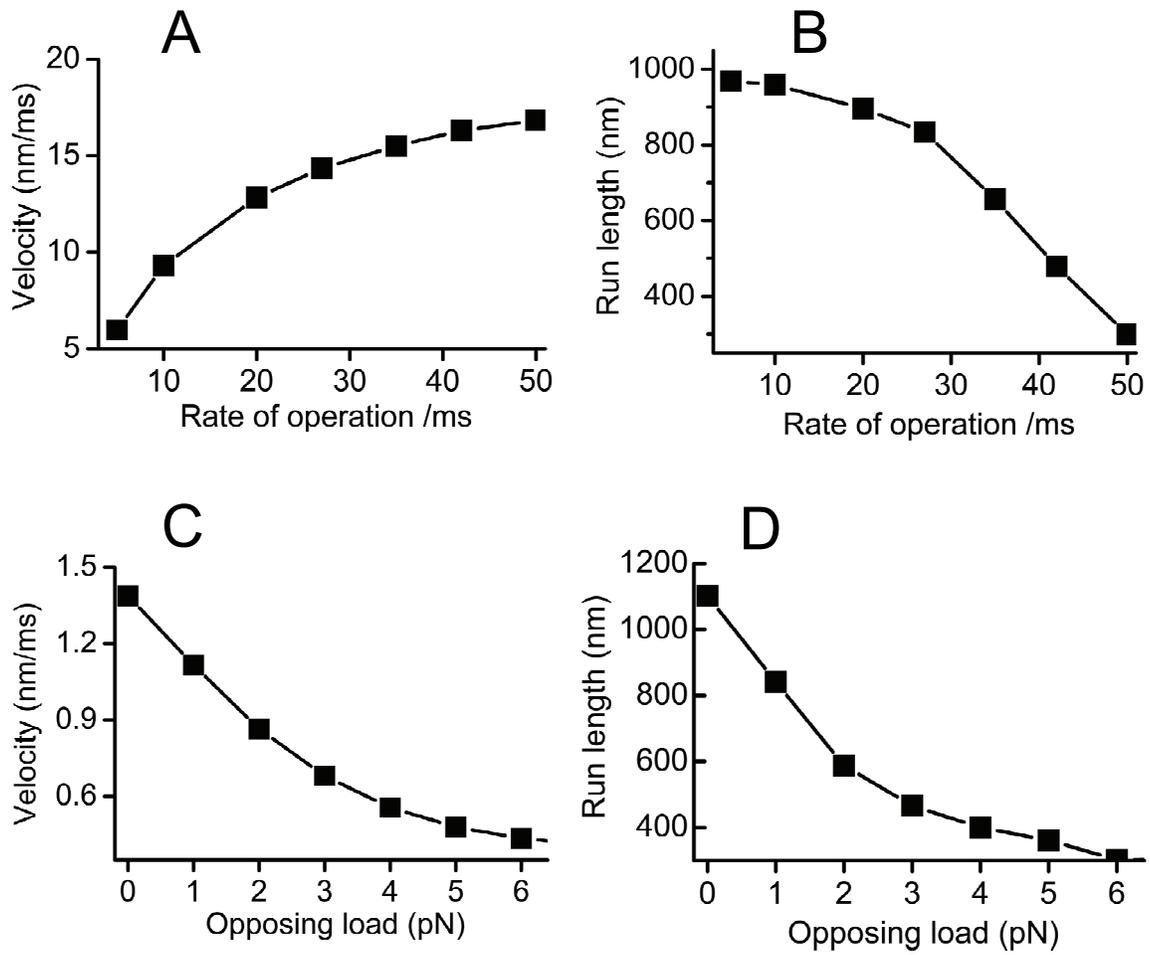

Figure 4

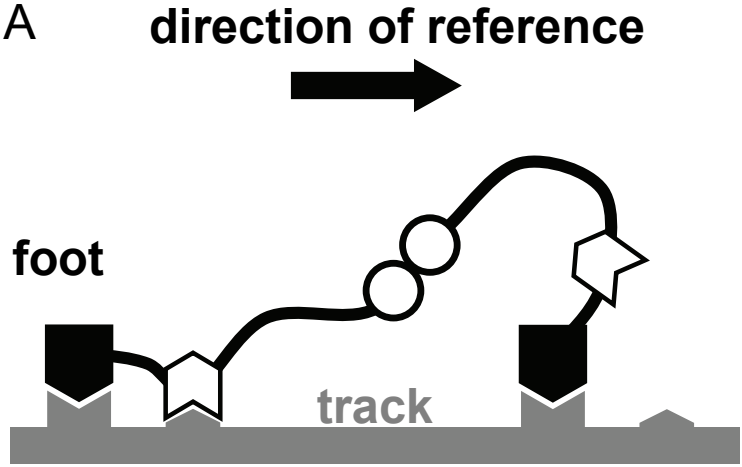

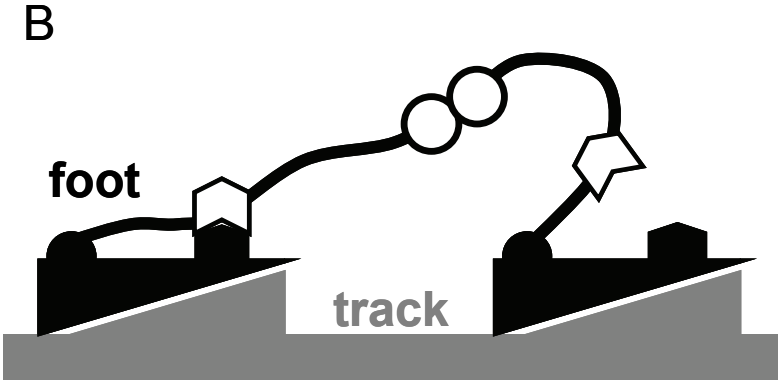